# *Effects of Selection Logging on Rainforest Productivity*

*Jerome K. Vanclay*

Queensland Forest Service, GPO Box 944, Brisbane 4000 Queensland, Australia

*Abstract*

An analysis of data from 212 permanent sample plots provided no evidence of any decline in rainforest productivity after three cycles of selection logging in the tropical rainforests of north Queensland. Relative productivity was determined as the difference between observed diameter increments and increments predicted from a diameter increment function which incorporated tree size, stand density and site quality. Analyses of variance and regression analyses revealed no significant decline in productivity after repeated harvesting. There is evidence to support the assertion that if any permanent productivity decline exists, it does not exceed six per cent per harvest.

## Introduction

Hilton (1987) claimed that rainforest logging could disrupt nutrient cycling and cause impoverishment, and argued that new harvesting methods were required. He suggested that rainfall made a considerable contribution to forest nutrition, and could replace the nutrients removed or lost during harvesting, provided that the residual stand was well stocked. He concluded that "so little hard information is available that no comparison can be made between past and present yields in those areas of rainforest which are being logged for the second time. Although the growing stock may have been accurately assessed before the first harvest the rate of growth remains unknown. Thus it is impossible to see ... how much productivity has been affected, if it has, by logging".

Enright (1978) reported that the growth rates of individual trees in the residual stand dropped markedly after logging. He also found that logging resulted in a temporary but marked decrease in nutrient levels after logging. However, both these studies concerned heavily logged stands. Enright (1978) reported that nearly all *Araucaria cunninghamii* (the dominant species comprising 54% of the stand) individuals exceeding 40 cm dbh were removed, and that extensive damage was caused to the residual stand.

Boxman *et al.* (1985) studied polycyclic logging followed by silvicultural treatment in Suriname and concluded that these contributed minimally to the loss of nutrients. Claims that polycyclic logging may lead to deterioration of the forest due to the progressive removal of the better genotypes have been refuted by Whitmore (1984), who argued that this was insignificant and academic.

The present study concerns the tropical rainforests of north-east Queensland. These forests had been managed for conservation and timber production for more than eighty years (Just 1987), before logging ceased following their World Heritage nomination in 1988. Although initial exploitation of these forests was largely uncontrolled, logging practices were progressively improved and harvesting in recent years has caused little environmental impact (Just 1987). The earliest exploitation caused relatively little damage because of the highly selective nature of logging and modest horsepower involved. Environmental impacts probably peaked during the mid-1960's with the ready availability of heavy earth moving machinery. During the 1980's, timber harvests were obtained through selection logging which removed 7 to 10 trees per hectare, comprising not more than 25 per cent of the total standing basal area (Vanclay 1989b). Guidelines (Preston and Vanclay 1988) ensured that not more than 50 per cent of the canopy was removed. Such guidelines ensure rapid recovery of the rainforest canopy (Horne and Gwalter 1982). Key components of this selection logging system as practiced during the 1980's were:

- Logging guidelines were sympathetic to the silvicultural requirements of the forest, viz. ensuring retention of vigorous advance growth, harvesting only defective and fully mature trees, providing for adequate regeneration of commercial species and discouraging invasion by weeds;
- Treemarking by trained staff specified trees to be retained, trees to be removed and the direction of felling to ensure minimal damage to growing stock and minimal opening of the canopy;
- Logging equipment was appropriate and driven by trained operators to ensure minimal damage to the residual stand and minimal soil disturbance, compaction and erosion;
- Prescriptions ensured that adequate stream buffers and steep slopes were excluded from logging;
- Sufficient areas for scientific reference, feature protection and recreation were identified and excluded from logging;
- Deficiencies in an evolving system were recognised and remedied, leading to an improved system.

Several studies have examined impacts of timber harvesting in these forests. Gilmour (1971) found that effects of logging on streamflow and sedimentation were small scale and short lived. Gillman *et al.* (1985) examined soil chemical properties and found that most topsoil nutrients regained their initial levels within four years of logging. Whilst nutrient cycles were disrupted by logging, losses appeared to be small and quickly replaced by natural inputs, provided that logging was of low intensity, short duration and infrequent (Congdon and Lamb 1990).

Nicholson *et al.* (1988, 1990) and Crome *et al.* (1990) reported that whilst timber harvesting caused localized destruction, it did not lead to loss of any plant species. Logging tracks and canopy loss were confined to 5 and 20 per cent of the area respectively (Crome *et al.* 1990). However, the light climate may be altered in areas with no direct canopy loss. Stocker (1981, 1983), Unwin (1983, 1988) and Webb and Tracey (1981) have investigated other aspects of the dynamics and regenerative capacity of these rainforests. Crome and Moore (1989, 1990) discussed effects of logging on fauna.

It has been estimated that a timber harvest of 60 000 cubic metres per annum could be sustained from these forests (Preston and Vanclay 1988). Vanclay and Preston (1989) examined the long-term sustainability of such a harvest, and concluded that selection logging could be sustained by the growth of residual trees and regeneration, and need not rely upon trees missed during previous harvests. Research into the relationship between diameter (breast high or above buttress, over bark) and log volume provided no evidence to suggest that there was any increase in defect or any reduction in log length in trees harvested from previously logged stands (Henry 1989).

Rainforests appear to have the regenerative capacity to cope with the effects of a single selection logging, given sufficient time to recover (Hopkins 1990). Shugart *et al.* (1980) used a succession model to examine the effects of comparatively intensive harvesting on a 30 year cycle in subtropical rainforest in New South Wales, and concluded that such harvesting was sustainable, although the structure and composition of the forest would be altered. The present paper examines how the long term average growth rates of individual trees are influenced by repeated selection logging.

## Data

The Queensland Department of Forestry (1983) research programme has provided an extensive database sampling virgin, logged and silviculturally treated forests. CSIRO (West *et al*. 1988) have also established 20 plots in relatively undisturbed stands sampling the full range of forest types in the region. The combined database represents over 250 permanent sample plots with a measurement history of up to 40 years (Appendix). Permanent sample plots range in size from 0.04 to 0.5 hectares, and have been re-measured frequently. All trees exceeding 10 cm dbh (diameter over bark at breast height or above buttressing) were uniquely identified and tagged, and were regularly measured for diameter (to nearest millimetre) using a girth tape. To improve the consistency of diameter measurement, field crews had access to previous records while in the field. Any trees exhibiting defects or bulges at or near the measurement height were noted and so identified on computer. Such trees have not been used in calculating diameter increments, and have only been used in calculating stand basal areas.

The data used in this study were identical to those used by Vanclay (1990) in developing a growth model for yield prediction (Vanclay and Preston 1989). Pairs of plot remeasurements were selected from the database to attain intervals between remeasurements of approximately five years, which did not span any logging or silvicultural activity. Tree diameters do not increase monotonically in size, but exhibit diurnal and seasonal fluctuations which may result in measured diameters smaller than previous values (Lieberman 1982). These, and measurement errors, may give rise to negative increments which may cause difficulties in data analysis. Ensuring a long interval between remeasurements (e.g. 5 years) so that the growth is large relative to the error, eliminates many of these decrements, but some remain. The logarithmic transformation used in the present analyses has long been recognised as an efficient way to satisfy assumptions implicit in regression analysis (linearity, normality, additivity and homogeneity of variance) (e.g. Schumacher 1939, Clutter 1963), but cannot accommodate negative increments. Some negative values can be accommodated by adding a constant before transforming, but any decrements exceeding 0.01 were omitted from the present analyses.

The data file created for statistical analysis contained 62 372 observations of diameter increment derived from 28 123 individual trees. The file also contained records of tree species and dbh, and stand variables such as site quality, stand basal area and soil parent material. Site quality for each plot was estimated using Vanclay's (1989a) Equation 13:

$$GI = \frac{\sum_{ij} Log(DI_{ij}+\alpha) - \sum_{ij}\left[\beta_{0i} + \beta_{1i}D_{ij} + \beta_{2i}Log(D_{ij}) + \beta_{3i}Log(BA) + \beta_{4i}OBA_{ij}\right]}{0.08808 \times \sum_{ij} Log(D_{ij})}$$

where *GI* is the growth index of the plot, $D_{ij}$ is the diameter (breast high or above buttress, over bark, in *cm*) of tree *j* of species *i*, *DI* is its diameter increment *(cm y$^{-1}$)*, $OBA_{ij}$ is its "overtopping basal area", the basal area of trees within the plot that are bigger than tree *ij* *(m$^2$ ha$^{-1}$)*, *BA* is the plot basal area *(m$^2$ ha$^{-1}$)*, and the *β*s are parameters estimated by linear regression. This equation estimates growth index, a measure of site productivity based on the diameter increment adjusted for tree size and competition, of all trees of eighteen reference species *(Acronychia acidula, Alphitonia whitei, Argyrodendron trifoliolatum, Cardwellia sublimis, Castanospora alphandii, Cryptocarya angulata, C. mackinnoniana, Darlingia darlingiana, Elaeocarpus largiflorens, Endiandra sp.* aff. *E. hypotephra, Flindersia bourjotiana, F. brayleyana, F. pimenteliana, Litsea leefeana, Sterculia laurifolia, Syzygium*

*kuranda, Toechima erythrocarpum, Xanthophyllum octandrum*) using all available remeasures for the plot (except that where plots were remeasured more frequently, remeasurements were selected to achieve approximately 5 year intervals). The *β*s were estimated by fitting the equation

$$Log (DI+ α) = Spp +D.Spp +Log (D).Spp +Log (BA).Spp +OBA.Spp +Log (D)Plot$$

(where *Spp* and *Plot* are qualitative variables) simultaneously for all these reference species in the development data set (80 plots, a further 64 plots were used for validation studies). The parameter α was assigned the value 0.02 after inspection of residuals and examining the residual mean squares from a range of values (Vanclay 1989a). The value 0.08808 was subjectively determined to scale the growth indices into the range 0-10.

Table 1. **Size and History of Plots used in Analyses.**

| Plot Area (ha) | Measurement History (Years) | | | | Total Plots |
|---|---|---|---|---|---|
| | 0.10 | 10-19 | 20-29 | 30+ | |
| <0.10 | 2 | 31 | 3 | 2 | 38 |
| 0.1-0.19 | 4 | 47 | 6 | 10 | 67 |
| 0.2-0.29 | 15 | 42 | 3 | 27 | 87 |
| ≥0.30 | 5 | 9 | 0 | 6 | 20 |
| Total | 26 | 129 | 12 | 45 | 212 |

The present study omitted any plots for which the estimated site quality exceeded the range 0-10, or for which the variance of the estimated site quality exceeded 2. Valid estimates of site quality were obtained for 212 plots (Table 1).

Table 2. **Logging History for Plot Data used in Analyses.**

| Harvests prior to First Measure | Total Harvests at Last Measure | | | | Total Plots |
|---|---|---|---|---|---|
| | 0 | 1 | 2 | 3 | |
| 0 | 9 | 1 | | | 10 |
| 1 | - | 136 | 38 | | 174 |
| 2 | - | - | 26 | 2 | 28 |
| Total | 9 | 137 | 64 | 2 | 212 |

Table 3. **Logging and Measurement Profile for Data used in Analyses**

| Year of Measure | Total Harvests prior to Measure | | | | Total Plots |
|---|---|---|---|---|---|
| | 0 | 1 | 2 | 3 | |
| -1949 | | 6 | 1 | | 7 |
| 1950-54 | 4 | 45 | 1 | | 50 |
| 1955-59 | 5 | 72 | 7 | | 84 |
| 1960-64 | 5 | 86 | 6 | | 97 |
| 1965-69 | 6 | 127 | 14 | | 147 |
| 1970-74 | 6 | 121 | 27 | | 154 |
| 1975-79 | 9 | 61 | 43 | 2 | 115 |
| 1980- | 4 | 23 | 21 | 2 | 50 |

Unfortunately, no continuous record of growth data spanning two successive harvests was available (Table 2). Experiment 615 (Appendix) had two such plots but the two-year period from establishment until logging was too short to provide reliable increment data. The 66 plots which were logged twice had measurement records which commenced only after the first harvest, and only one plot had a measurement record spanning the first harvest. However,

38 plots provided measurement data spanning the second harvest (Table 2). Only two plots were logged three times, but these harvests differed from normal practice in that the first and second harvests were only seven years apart (Appendix, Experiment 615). The majority of plots (136) were in stands logged once, and had a measurement record which did not span any logging activity.

A further problem was that these experiments were not well replicated through time - most were first logged during the decade 1950-59, and were relogged during 1969-80 (Appendix). Table 3 shows that the different harvesting histories were well sampled, but these plots were not necessarily paired with suitable control plots. Thus differences detected during any given period could be due to site or management differences, as well as to harvesting history. Similarly, on any given plot, differences in increment between measurement periods could be due to prevailing weather conditions, as well as due to harvesting. Thus although extensive, the present database contained weaknesses which provided problems for the analysis and interpretation of the results.

The severity of these problems may be gauged through a correlation matrix of plot variables (Table 4). Ideally, the explanatory variables explored in an analysis should not be correlated, although in practice this is rarely possible. When explanatory variables are correlated, the ability to identify potentially causal relationships is reduced (explanatory variables do not have a unique sum of squares), and the magnitude of possible effects may not be able to be reliably determined (addition or subtraction of an explanatory variable may substantially change parameter estimates for a model, standard errors of estimates may be inflated). However, multicollinearity does not inhibit the ability to obtain a good fit, nor does it affect inferences about responses or predictions within the region of observations (Neter and Wasserman 1974:341). Correlations between explanatory variables considered in the present analysis are not serious (Table 4). The high correlation between number of harvests and time since logging (-0.64) is partly due to the encoding convention adopted (for unlogged plots, time since logging = 99), and the correlation for logged plots is lower (-0.43).

## Hypothesis and Analyses

The analyses test the hypothesis that selection logging leads to a reduction in productivity in these rainforests and that this reduction may comprise two components, a transient and a permanent loss of productivity. Figure 1 illustrates the pattern of productivity decline that the analyses attempt to detect. The null hypothesis was that there is no reduction in productivity, whilst the alternative hypothesis was that a reduction in productivity following logging can be detected. The analyses endeavour to produce evidence to reject the null hypothesis.

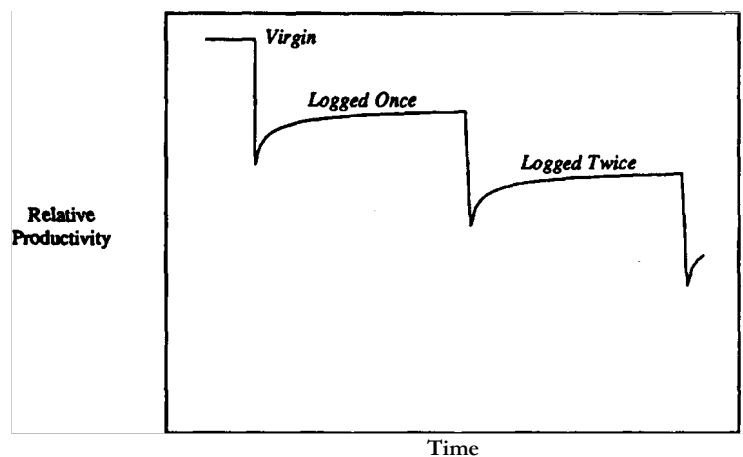

**Figure 1. Hypothetical effect of logging on productivity.**

**Table 4. Correlation Matrix for Plot Variables:**

| Source | No of harvests | Time since treatment | Time since logging | Site quality | Basal Area | Mean Residual |
|---|---|---|---|---|---|---|
| Year of measure | 0.272* | 0.071 | 0.088* | -0.019 | 0.040 | -0.092* |
| No of harvests | 1.000 | -0.252* | -0.639* | -0.084* | -0.230* | -0.079* |
| Time since treatment |  | 1.000 | 0.473* | 0.326* | 0.741* | -0.055 |
| Time since logging |  |  | 1.000 | 0.207* | 0.588* | 0.019 |
| Site quality |  |  |  | 1.000 | 0.615* | 0.006 |
| Stand basal area |  |  |  |  | 1.000 | 0.101* |

\* indicates correlation significant at P<0.05

Unfortunately, a suitable measure of "productivity" is neither easy to define nor to measure. Biomass production may seem a good measure of productivity, but has several weaknesses. It cannot be measured directly, and is difficult to determine. Nett biomass production is near zero in unlogged stands (any growth is offset by mortality) and increases following logging due in part to a reduction in competition. Gross biomass production overcomes the problem of mortality, but is dependent upon stocking, and a reduction in production following logging could be due to the reduced occupancy of the site. This problem of distinguishing the effects of site occupancy from the effects of logging is common to all stand level measures, including volume and basal area increment per hectare. Thus we need to consider individual trees, and could seek to monitor the growth of a "standard reference tree" in each plot. However, suitable trees having the same species, size and competition do not exist in each plot. Even if such trees could be found in a number of plots, logging would reduce competition and bias comparisons with unlogged plots. One solution to this dilemma is to fit a regression equation to the individual tree increments, and examine the residuals obtained from comparing the observed and expected increments. This approach is widely used in many disciplines, most commonly to derive seasonally adjusted figures (e.g. below average temperatures for June take into account that it is winter; seasonally adjusted employment figures account for school-leavers in December). Keenan and Candy (1983) used residuals about a height-age curve to investigate site factors influencing *Eucalyptus delegatensis* regrowth.

Suitable residuals can be generated from published increment equations. Vanclay (1990) presented 41 equations to predict the diameter increment of the 400 species occurring in the database. These equations had the form:

$$Log(DI+0.02) = \beta_0 + \beta_1 \times D + \beta_2 \times Log(D) + \beta_3 \times Log(D) \times SQ + \beta_4 \times Log(BA) \quad (1)$$
$$+ \beta_5 \times OBA + \beta_6 \times PS + \beta_7 \times TST \times e^{-TST/5}$$

where *DI* is diameter increment *(cm y$^{-1}$)*, *D* is dbh *(cm)*, SQ is site quality (Vanclay 1989a), *BA* is stand basal area *(m$^2$ ha$^{-1}$)* of trees exceeding 10 cm dbh, *OBA* is overtopping basal area *(m$^2$ ha$^{-1}$)*, defined as the basal area of stems whose diameter exceeds that of the subject tree, *TST* is time *(years)* since silvicultural treatment, PS is a binary variable which takes the value one if the species is growing on a "preferred soil parent material" and zero otherwise, and the *β*s are parameters specific to each species group.

This equation does not include expressions of the number of harvests or of the time since logging. Thus the residuals should indicate the effects of logging and other factors not considered in Equation (1). Figure 2 illustrates these residuals (representing means of 8000, 4000, 3000 and 450 tree remeasurements for 0, 1, 2 and 3 harvests respectively). These suggest some productivity change with time since logging, but little effect attributable to number of harvests and little resemblance to Figure 1.

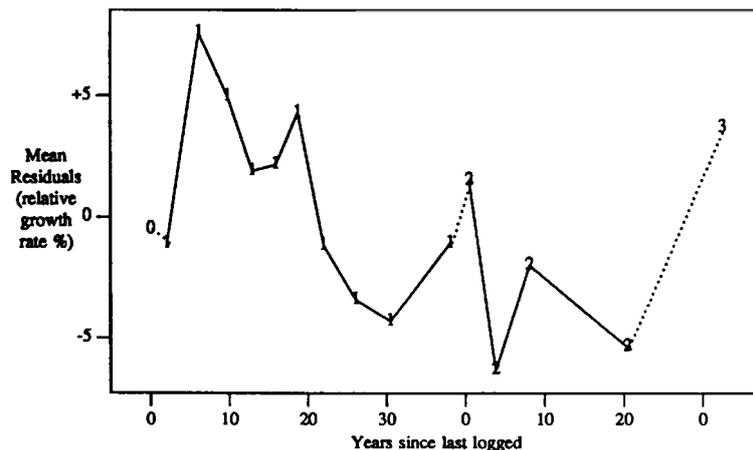

**Figure 2. Difference between observed and predicted diameter increments for virgin stands (0) and stands logged once (1), twice (2) or three times (3).**

An analysis of variance of the residuals about Equation (1) enables a formal statistical test of the hypothesis to be made. However, such an analysis of variance can be conducted in several different ways, and can test several different factors. In compiling the analysis then, it is essential to take account of the particular characteristics of the present data. In particular, we have data from 62 372 remeasurements on individual trees to infer the effects of logging on 212 plots. These individual tree data may give an inflated estimate of precision and may place undue emphasis on well stocked plots, so it is appropriate to calculate the mean residual for each plot remeasurement and use that in further analyses. Not all plot remeasurements give rise to a mean residual of equal precision, so it is appropriate to weight the analysis by the inverse of the variance associated with the mean residual for each plot remeasurement. Such weighting ensures that those plots which exhibit the most consistent growth patterns have greater influence on the analysis. Some plot remeasurements exhibited very small variances which would have given rise to inappropriately large weights. Thus 12 data with small variances were assigned the value 0.1. Weights were adjusted so that the sum of the weights equalled the number of data, and the final weights ranged from 0.1 to 4.2.

Table 5 reports several factors examined in an analysis of variance. Time since last logging was represented as six intervals of five years (0-4, 5-9, ..., 25+ years), and other periodic effects were taken into account through several approximately five year intervals (pre-1955, 1955-59, ..., 1980+). This analysis revealed that soil parent material, period of observation and the interaction between soil and time since logging were significant ($P < 0.05$) in influencing the differences between observed and expected diameter increments. The significant factors could be due to management practices as well as to environmental effects. Soil parent material influences topographic slope as well as soil type, and slope is a major determinant of logging damage (Vanclay 1989b). A problem with multicollinear data is that the explanatory variables do not have a unique sum of squares (Neter and Wasserman 1974:341), and that the significance associated with a variable may depend on the order in which the variables were included in the model. One way to overcome this is to determine the sum of squares for each variable by subtracting it from the maximal model. Whilst this ensures unique sums of squares for each variable, an additional entry in the analysis of variance table is required to reconcile the sums of squares (e.g. Table 5). This entry also indicates the extent of multicollinearity.

Table 6 reports the changes in productivity estimated through the analysis of variance. None of the estimates in Table 6 differ significantly from zero, and there is no suggestion of productivity decline. Table 6 does not aid in the detection of long-term decline, as the fluctuating response suggested does not enable forecasts. To detect a long term trend, we need to reformulate the model with number of harvests as a linear variate rather than a factor. This has the effect of estimating an equal and cumulative change in productivity following each successive harvest, as is illustrated in Figure 1. Two linear transformations of time since logging were also explored. One option is to use the inverse of time since logging, implying the asymptotic trend illustrated in Figure 1. Another option is to use a transformation similar to that used for the response to silvicultural treatment ($te^{-t/\alpha}$) which predicts a maximum response in year a followed by an asymptotic return to zero. The present data support the latter transformation $(te^{-t})$ with a very short-lived response ($\alpha = 1$). This linear transformation provided a better fit with fewer degrees of freedom than the inclusion of time since logging as a factor. However, the present data were derived from measurements over approximately 5-year intervals and are not suited for determining the exact nature of this short-term response. Including the number of harvests as a linear variate rather than a factor led to a slight increase in the residual sum of squares (P = 0.13). The analysis of variance (Table 7) was not greatly affected by the use of linear variates; time since logging was significant (P < 0.05) as a linear variate whilst number of harvests remained non-significant (P > 0.05).

Table 5. Analysis of Variance of Mean Plot Residuals using four Factors with Interactions.

| Source of Variation | Degrees of Freedom | Residual Sum of Squares | Residual Mean Square | Test Statistic: F-ratio | Probability | Significance |
|---|---|---|---|---|---|---|
| MAIN EFFECTS | | | | | | |
| Soil parent material | 5 | 1.920 | 0.3840 | 4.69 | 0.0005 | *** |
| 5-year period | 6 | 1.357 | 0.2262 | 2.76 | 0.012 | * |
| No. of harvests | 3 | 0.411 | 0.1369 | 1.67 | 0.2 | |
| Time since logging | 5 | 0.405 | 0.0811 | 0.99 | 0.6 | |
| Multicollinearity | 0 | 0.981 | | | | |
| INTERACTIONS | | | | | | |
| Soil-Time | 15 | 2.470 | 0.1647 | 2.01 | 0.013 | * |
| Soil-Harvests | 4 | 0.343 | 0.0858 | 1.05 | 0.4 | |
| Period Time | 27 | 2.196 | 0.0813 | 0.99 | 0.5 | |
| Period-Harvests | 13 | 1.020 | 0.0785 | 0.96 | 0.5 | |
| Period-Soil | 18 | 1.325 | 0.0736 | 0.90 | 0.6 | |
| Harvests-Time | 5 | 0.304 | 0.0608 | 0.74 | 0.6 | |
| Multicollinearity | 6 | 0.3671 | | | | |
| Residual | **65** | 53.967 | 0.0819 | | | |
| Total | 76̂6 | 67.066 | | | | |

Table 6. Productivity of Logged Forest relative to Virgin Forest..

| Factors Considered | Number of Harvests | | |
|---|---|---|---|
| Number of Harvests only, significant terms omitted | +3% | -5% | +4% |
| All Main Effects except time since logging | *+5%* | -3% | +20% |
| All Main Effects including time since logging | +8% | +2% | +27% |
| All Main Effects & Significant Interactions | +7% | +5% | +16% |

N.B. None of these estimates is significantly different from zero.

A number of alternative approaches can be used to explore the proposed hypothesis. One alternative is to perform an analysis of variance on the individual tree data, ignoring the implications of the inflated degrees of freedom and unequal weighting of plots (Table 8, Model 2). Another possibility is to fit Equation (1) simultaneously to all 41 species groups, and to include additional variables for number of harvests, time since logging, time period and soil parent material (Table 8, Model **3).** Both these approaches indicated that both number of harvests and time since logging were not significant (P > 0.2), whilst five-year period and soil parent material remained significant (P < 0.001). .

Yet another technique is to identify comparable plots with different logging histories, and to eliminate other factors which may confound the result. Selection of plots was based on several objective criteria:

- Plots at least 0.2 hectares in area
- Established prior to 1960
- Maintained at least until 1980
- Measurement history spanning at least 30 years, and
- Plot site quality determined with variance not exceeding 0.1.

The majority of the plots satisfying these criteria were located on soils derived from coarse-grained granites, so selection was further restricted to the 16 plots with this soil parent material. This selection included (see Appendix) Experiments 591, 612, 613 (Plots 1& 3), 615, 616 and 619 (Plot 1). Analysis of variance and regression analysis indicated that none of the factors considered (number of harvests, time since logging, 5-year period) were significant (P > 0.05).

**Table 7. Analysis of Variance of Mean Plot Residuals using Linear Variates.**

| Source of Variation | Degrees of Freedom | Sum of Squares | Mean Square | Test Statistic: F-ratio | Probability | Significance |
|---|---|---|---|---|---|---|
| 5-year period | 6 | 1.923 | 0.321 | 3.87 | 0.0011 | ** |
| Soil parent material | 5 | 1.740 | 0.348 | 4.20 | 0.0012 | ** |
| Time since last logged | 1 | 0.524 | 0.524 | 6.32 | 0.012 | * |
| No. of harvests | 1 | 0.021 | 0.021 | 0.25 | 0.6 | |
| Multicollinearity | 0 | 0.502 | | | | |
| Residual | 753 | 62.356 | 0.083 | | | |
| Total | 766 | 67.066 | | | | |

The parameter estimates given in Table 8 enable the effects of time since logging and logging history to be assessed. The test statistic (student's t) indicates the statistical significance of the response, and parameter estimates enable the effect of logging to be quantified. For example, the mean plot residuals (Table 8, Model 1) give rise to a parameter estimate of -0.01308 for number of harvests which suggests that productivity will decrease relative to the unlogged condition to $e^{-0.01308} = 0.987$ after the first harvest, to 0.974 after the second, etc. Similarly, the transient response (time since logging) predicts a decrease of $e^{-0.3117t/e^t} = 0.892$ in the first year after logging, 0.919 in the second year, and 0.990 in year 5.

**Table 8. Parameter Estimates and Implied Productivity Change due to Selection Logging.**

| Model | Number of Harvests (Permanent Change) | | | | Time since Logging (Transient Change) | | | |
|---|---|---|---|---|---|---|---|---|
| | Parameter Estimate | Standard Error | Student's t | Implied Change | Parameter Estimate | Standard Error | Student's t | Implied Change |
| All available data: | | | | | | | | |
| 1) Mean Residuals | -0.01308 | 0.02592 | 0.505 | -1% | -0.3117 | 0.1239 | 2.516* | -11% |
| 2) Individual Trees | -0.00111 | 0.00597 | 0.186 | 0% | -0.0649 | 0.0390 | 1.664 | -2% |
| 3) Expanded Model | +0.00165 | 0.00712 | 0.232 | 0% | -0.0556 | 0.0406 | 1.369 | -2% |
| Subset only: | | | | | | | | |
| 4) Mean Residuals | -0.00856 | 0.03251 | 0.263 | -1% | +0.0288 | 0.2289 | 0.126 | +1% |
| 5) Individual Trees | +0.01108 | 0.01085 | 1.021 | +1% | +0.1050 | 0.0918 | 1.144 | +4% |

* indicates significantly different from zero at P< 0.05

## Discussion

The analysis of variance of mean plot residuals reported in Table 7 provides no evidence to reject the null hypothesis that harvesting causes no permanent decline in productivity. Model 1 provides evidence to support the existence of a transient decline in productivity during the few years following logging. Other analyses of individual tree data, and of the selected subset of data (Table 8) provided no evidence to reject the null hypothesis.

The parameter estimates for the terms reflecting the permanent impact of harvesting (number of harvests) are always close to zero, never significantly different from zero, and do not differ significantly from one another despite their different signs (Table 8). Because these parameters are very close to zero, one should not place too much emphasis on this change of sign, but it may be attributed, in part, to the non-orthogonal nature of the data. This weakness is inevitable in opportunistic analyses of this sort, and can only be overcome by properly designed and replicated experiments. Unfortunately, the large areas of virgin rainforest and long time period required to conduct such a properly designed experiment probably render such experimental results unattainable. In any case, such an experiment would not yield useful results for several decades, so the present data provide the only means currently available to assess the long term effects of repeated logging. Fortunately, multicollinearity does not inhibit our ability to develop a good model from the data, or to make inferences from that model (Neter and Wasserman 1974:341).

The parameter estimate for the transient decline in productivity (time since logging) from Model 1 differs significantly those obtained from other approaches. This analyses of plot mean residuals (Model 1) found a significant but short-lived transient decline in productivity. It is beyond the scope of the present study to determine possible causes: it may be that logging created an environment less favourable for growth; it may equally well be that trees were directing photosynthates into canopy expansion rather than diameter increment. This significant transient decline in productivity was not detected in Models 2-5, the parameter estimates of which did not differ significantly from zero or from one another.

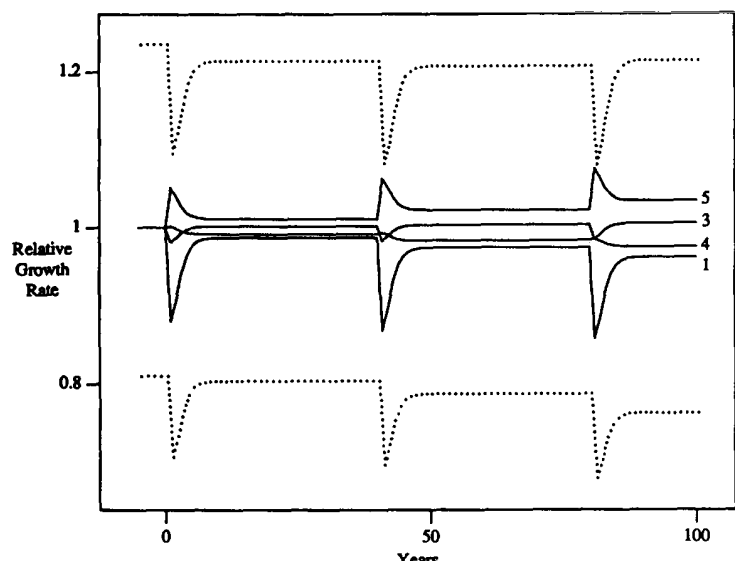

**Figure 3. Predicted effects of logging on productivity. Models are indicated by numbers. Dotted lines represent 95% confidence limits.**

The implications of Models 1, 3, 4 and 5 of Table 8 are illustrated in Figure 3. Approximate 95% confidence intervals for Model 1 are also shown, and indicate that the illustrated models do not differ significantly from the unlogged condition. Thus the principle of parsimony leads us to accept the null hypothesis that logging has no permanent effect on productivity. There is no evidence to reject the null hypothesis and support the alternate hypothesis, as any apparent change in productivity is not significant and may be due to random variation.

Which is the most appropriate model? This question is largely academic as no model supports the existence of a permanent decline in productivity. As previously argued, Models 1 and 4 which examine the plot mean residuals, are attractive. Model 3 estimates the logging effects directly from the raw data rather from partial residuals and may be less subject to effects multicollinearity, but has inflated degrees of freedom and may underestimate the standard errors.

Tests so far have adopted the conventional parsimonious approach of accepting the null hypothesis (that logging causes no decline in productivity) unless there is strong evidence to the contrary. However, statistical tests can also be formulated to test the null hypothesis that logging causes an x per cent decline in productivity, and to reject this only if there is strong evidence that any decline is less than this specified amount. This places the burden of proof on the forest manager. Suppose we have reason to suspect that each harvest causes a permanent and cumulative five per cent decline in productivity (this assumes a parameter estimate of -0.05130 for number of harvests). The data tabulated in Table 9 provide evidence to reject this contention ($P < 0.1$ for Models 1 & 4, $P < 0.0001$ for Models 2, 3 & 5). It is also interesting to examine the critical values x which would just lead to the rejection of the null hypothesis that logging led to a decline in productivity of (or exceeding) x per cent (Table 9). The models based on plot mean residuals (Models 1 & 4) are cautious models and allow the possibility of a small productivity decline not admitted by models based on individual tree data (Models 2, 3 & 5). This may be attributed in part to the inflated degrees of freedom and underestimation of standard errors in the individual tree models (2, 3 & 5). However, the multicollinearity evident in the data (Table 4) would lead to inflated estimates of standard error (Neter and Wasserman 1974:341) with the result that the critical values for Models 1 and 4 (Table 9) may be unnecessarily cautious.

**Table 9. Critical Values for Rejection of the Hypothesis that Logging Causes Productivity Decline.**

| Model | Critical Values to Reject Permanent Decline | | | Probability of Permanent Decline | |
|---|---|---|---|---|---|
| | P=0.05 | P=0.01 | P=0.01 | 2% | 5% |
| 1 | -5.4% | -7.1% | -9.1% | 0.4 | 0.07 |
| 2 | -1.1% | -1.5% | -2.0% | 0.001 | < 0.0001 |
| 3 | -1.0% | -1.5% | -2.1% | 0.0013 | < 0.0001 |
| 4 | -6.0% | -8.1% | -10.6% | 0.4 | 0.09 |
| 5 | -0.7% | -1.4% | -2.3% | 0.002 | < 0.0001 |

It is appropriate to observe that no evidence exists of any long-term decline in productivity following repeated harvesting. Whilst there is insufficient evidence to reject the possibility of a small decline, there is evidence to support the assertion that any decline does not exceed six per cent per harvest.

## Conclusion

These analyses reveal no evidence to suggest any long-term decline in rainforest productivity after three cycles of selection logging. Despite an extensive database incorporating over 200 plots, some established more than 40 years, the data are inadequate for conclusive studies on the long term effects of rainforest harvesting. Continued monitoring and additional harvesting of experimental plots will be necessary to conclusively demonstrate the long term effects of logging.

However, the present analyses provide no evidence of any long-term decline in productivity following three cycles of conservative polycyclic selection logging, and provide evidence that any decline does not exceed six per cent per harvest. These results should not be extrapolated to infer the sustainability of more intensive harvesting systems.

**Acknowledgements**
Many officers of the Queensland Department of Forestry have contributed to the establishment and maintenance of experiments and the database. It is due to their diligence that analyses such as these are possible. Special thanks are due to Greg Unwin for providing data from the CSIRO EP series of plots, and to John Rudder for assistance with data processing. Drs. H. C. Dawkins, D. Doley, D. Lamb and an anonymous referee provided helpful comment on an earlier manuscript.

## Appendix

The following is a list of permanent sample plots in Queensland Forest Service rainforest database at the time this study was commenced. Only those plots with a valid site quality were used in the present analyses. Geological types are Alluvial (AL), Acid Volcanic (AC), Basic Volcanic (BV), Coarse-grained Granite (CG), Sedimentary and Metamorphic (SM) and Tully fine-grained Granite (TG). Rainforest structural types follow Tracey and Webb (1976). Brief descriptions of the origin of the various plot types are given below.

| Expt No | Plot No | State Forest | AMG Grid Ref. | Area (ha) | First measure | Last measure | Geol type | Site qual | Alt. (m) | Aspect | Slope (deg) | Rain (mm) | Struct type | Years logged | Years treated | Plot type |
|---|---|---|---|---|---|---|---|---|---|---|---|---|---|---|---|---|
| 69 | 1 | 185 | 55 349700 8101050 | 0.4047 | 48 | 59 | SM | 5.0 | 670 | WSW | 5 | 1320 | 6 | | | 1 |
| 77 | 1 | 185 | 55 348800 8101300 | 0.4047 | 48 | 57 | SM | 6.0 | 670 | N | 10 | 1320 | 6 | 43 | | 1 |
| 77 | 2 | 185 | 55 348860 8101300 | 0.4047 | 48 | 57 | SM | 7.4 | 670 | N | 10 | 1320 | 6 | 43,52 | 52 | 1 |
| 78 | 1 | 185 | 55 349690 8100330 | 0.4047 | 48 | 87 | BV | 7.8 | 680 | NNW | 5 | 1320 | 6 | 43 | | 1 |
| 78 | 2 | 185 | 55 349690 8100290 | 0.4047 | 48 | 87 | SM | 2.0 | 680 | NNW | 5 | 1320 | 6 | 43,49 | 49 | 1 |
| 79 | 1 | 185 | 55 349100 8101300 | 0.4047 | 49 | 57 | SM | 4.9 | 670 | N | 10 | 1320 | 6 | 43 | 51 | 1 |
| 79 | 2 | 185 | 55 349160 8101300 | 0.4047 | 49 | 57 | SM | 5.5 | 670 | N | 10 | 1320 | 6 | 43 | 51 | 1 |
| 89 | 1 | 191 | 55 340090 8082800 | 0.0405 | 51 | 64 | BV | 9.0 | 680 | - | 0 | 1400 | | 27 | 51 | 2 |
| 89 | 2 | 191 | 55 340090 8082780 | 0.0405 | 53 | 64 | B V | - | 680 | | 0 | 1400 | | 27 | 51 | 2 |
| 99 | 1 | 191 | 55 339030 8082560 | 0.1036 | 52 | 70 | BV | - | 680 | | 0 | 1400 | | 28 | 53 | 2 |
| 99 | 2 | 191 | 55 339030 8082560 | 0.1036 | 52 | 70 | BV | - | 680 | SE | 5 | 1400 | | 28 | 53 | 2 |
| 99 | 3 | 191 | 55 339030 8082560 | 0.1012 | 52 | 70 | BV | - | 680 | | 0 | 1400 | | 28 | 53 | 2 |
| 99 | 4 | 191 | 55 339030 8082560 | 0.1012 | 61 | 73 | BV | - | 680 | | 0 | 1400 | | 28 | 53 | 2 |
| 99 | 5 | 191 | 55 339030 8082560 | 0.0838 | 52 | 70 | BV | - | 680 | SE | 5 | 1400 | | 28 | 53 | 2 |
| 99 | 6 | 191 | 55 341100 8082510 | 0.0979 | 52 | 87 | BV | - | 680 | | 0 | 1400 | | 28 | 52 | 2 |
| 99 | 7 | 191 | 55 341190 8082580 | 0.1024 | 52 | 87 | BV | - | 680 | | 0 | 1400 | | 28 | 52 | 2 |
| 110 | 2 | 310 | 55 361090 8086740 | 0.1012 | 52 | 68 | BV | 5.6 | 670 | N | 5 | 2000 | | 30,68 | 30,53 | 3 |
| 111 | 1 | 185 | 55 350120 8099160 | 0.1578 | 52 | 68 | SM | 5.5 | 680 | N | 10 | 1320 | 6 | 39 | | 4 |
| 111 | 2 | 185 | 55 350020 8099130 | 0.1348 | 52 | 68 | SM | - | 670 | W | 10 | 1320 | 6 | 39 | 52 | 4 |
| 111 | 3 | 185 | 55 350200 8099090 | 0.1643 | 52 | 68 | SM | 7.6 | 680 | SE | 10 | 1320 | | 39 | 52 | 4 |
| 137 | 1 | 194 | 55 331410 8086410 | 0.1060 | 54 | 77 | CG | 4.6 | 1080 | W | 5 | 1650 | | 53,80 | 53,57 | 5 |
| 159 | 1 | 191 | 55 339670 8082920 | 0.1012 | 54 | 70 | BV | - | 680 | | 0 | 1400 | 5b | 33 | 54,62 | 5 |
| 159 | 2 | 191 | 55 339580 8082900 | 0.1012 | 54 | 70 | BV | 1.4 | 680 | | 0 | 1400 | 5b | 33 | 54,62 | 5 |
| 159 | 3 | 191 | 55 339580 8082990 | 0.1012 | 54 | 70 | BV | 5.0 | 680 | | 0 | 1400 | 5b | 33 | 54,62 | 5 |
| 159 | 4 | 191 | 55 339660 8083000 | 0.1012 | 54 | 70 | BV | 3.0 | 680 | | 0 | 1400 | 5b | 33 | 54,62 | 5 |
| 159 | 5 | 191 | 55 339650 8083080 | 0.1012 | 55 | 70 | BV | 4.6 | 680 | | 0 | 1400 | 5b | 33 | 54,58,62 | 5 |
| 159 | 6 | 191 | 55 339610 8083070 | 0.1012 | 55 | 70 | BV | 2.0 | 680 | - | 0 | 1400 | 5b | 33 | 54,58,62 | 5 |
| 159 | 7 | 191 | 55 339570 8083060 | 0.1012 | 55 | 70 | BV | - | 680 | - | 0 | 1400 | 5b | 40 | 54,58,62 | 5 |
| 166 | 1 | 251 | 55 347980 8038080 | 0.4047 | 69 | 83 | BV | 7.0 | 720 | W | 10 | 1800 | | 55 | 56,57,62 | 5 |
| 166 | 2 | 251 | 55 347960 8038060 | 0.4047 | 69 | 83 | BV | 7.1 | 720 | W | 10 | 1800 | | 55 | 56,57,62 | 5 |
| 167 | 1 | 194 | 55 331660 8086520 | 0.3541 | 54 | 63 | CG | 4.0 | 1060 | - | 0 | 1650 | | 53 | 54 | 5 |
| 167 | 2 | 194 | 55 331640 8086570 | 0.2023 | 55 | 63 | CG | 7.2 | 1060 | - | 0 | 1650 | | 53 | 54 | 5 |
| 174 | 1 | 310 | 55 361450 8086720 | 0.1010 | 54 | 68 | BV | 5.3 | 670 | SSW | 5 | 2000 | | 29,68 | 29,54 | 3 |
| 174 | 2 | 310 | 55 361450 8086720 | 0.1008 | 54 | 68 | BV | 5.0 | 670 | ESE | 5 | 2000 | | 29,68 | 29,54 | 3 |
| 178 | 1 | 1229 | 55 350960 8146940 | 0.0283 | 55 | 62 | SM | 0.2 | 440 | - | 0 | 2090 | 12c | 50,78 | 55,62 | 5 |
| 178 | 2 | 1229 | 55 350980 8146940 | 0.0809 | 55 | 76 | SM | 3.2 | 440 | N | 5 | 2090 | 12c | 50,78 | 55,62 | 5 |

| | | | | | | | | | | | | | | | | | |
|---|---|---|---|---|---|---|---|---|---|---|---|---|---|---|---|---|---|
| 178 | 3 | 1229 | 55 351060 8146940 | 0.0348 | 55 | 62 | SM | 3.7 | 440 | - | 0 | 2090 | 12c | 50,78 | 55,62 | 5 |
| 180 | 1 | 1229 | 55 351300 8147000 | 0.0769 | 56 | 78 | SM | 5.7 | 440 | - | 0 | 2030 | 2a | 51,77 | 56 | 5 |
| 184 | 1 | 310 | 55 357970 8090050 | 0.4047 | 55 | 68 | BV | 4.7 | 720 | - | 0 | 2030 | 1b | 58 | 59 | 1 |
| 184 | 2 | 310 | 55 358100 8089950 | 0.4047 | 55 | 68 | BV | 6.2 | 720 | - | 0 | 2030 | 1b | 58 | 59 | 1 |
| 207 | 1 | 194 | 55 332850 8087000 | 0.1012 | 56 | 68 | CG | 8.8 | 1130 | N | 5 | 1650 | 9 | 68 | | 6 |
| 222 | 1 | 310 | 55 361180 8086730 | 0.1036 | 58 | 68 | BV | 6.2 | 670 | NNW | 5 | 2000 | | 28 | 29 | 3 |
| 222 | 2 | 310 | 55 361130 8086680 | 0.1012 | 58 | 68 | BV | 5.8 | 670 | NNW | 5 | 2000 | | 28 | 29,58 | 3 |
| 222 | 3 | 310 | 55 361060 8086610 | 0.1012 | 58 | 68 | BV | 5.8 | 670 | NNW | 5 | 2000 | | 28 | 29,58 | 3 |
| 222 | 4 | 310 | 55 361110 8086680 | 0.1004 | 58 | 68 | BV | 5.9 | 670 | NNW | 10 | 2000 | | 28 | 29,58 | 3 |
| 224 | 1 | 310 | 55 361470 8086320 | 0.1012 | 58 | 68 | BV | 7.4 | 670 | SW | 10 | 2000 | | 28 | 29 | 3 |
| 224 | 2 | 310 | 55 361470 8086390 | 0.1000 | 58 | 68 | BV | 8.0 | 670 | SW | 10 | 2000 | | 28 | 29,58 | 3 |
| 224 | 3 | 310 | 55 361560 8086360 | 0.1012 | 58 | 68 | BV | 6.1 | 670 | SW | 5 | 2000 | | 28 | 29,58 | 3 |
| 224 | 4 | 310 | 55 361580 8086360 | 0.1012 | 58 | 68 | BV | 5.9 | 670 | SW | 5 | 2000 | | 28 | 29,58 | 3 |
| 224 | 5 | 310 | 55 361330 8086390 | 0.1008 | 58 | 68 | BV | 6.0 | 670 | NE | 5 | 2000 | | 28 | 29,58 | 3 |
| 224 | 6 | 310 | 55 361330 8086440 | 0.1020 | 58 | 68 | BV | 7.1 | 670 | NE | 5 | 2000 | | 28 | 29,53 | 3 |
| 226 | 1 | 310 | 55 358520 8090320 | 0.0777 | 58 | 74 | BV | - | 670 | NNE | 5 | 1800 | | 57 | 58 | 5 |
| 241 | 1 | 310 | 55 358600 8090400 | 0.1267 | 59 | 75 | BV | 6.4 | 720 | NE | 25 | 2030 | 1b | 58 | 59 | 5 |
| 242 | 1 | 194 | 55 331700 8084050 | 0.1117 | 59 | 74 | AV | 4.6 | 1035 | N | 25 | 1650 | 9 | 58 | 58 | 5 |
| 243 | 1 | 194 | 55 332870 8089170 | 0.2598 | 59 | 74 | CG | 3.8 | 980 | W | 10 | 1650 | | 54 | 56,59 | 5 |
| 245 | 1 | 1229 | 55 349910 8147220 | 0.2068 | 59 | 72 | SM | 3.2 | 440 | - | 0 | 2030 | 2a | 58 | 59 | 5 |
| 245 | 2 | 1229 | 55 349880 8147260 | 0.2262 | 59 | 72 | SM | 4.2 | 440 | - | 0 | 2030 | 2a | 58 | 59 | 5 |
| 245 | 3 | 1229 | 55 349940 8147260 | 0.1941 | 59 | 72 | SM | 2.3 | 440 | - | 0 | 2030 | 2a | 58 | 59 | 5 |
| 246 | 1 | 1229 | 55 351480 8146450 | 0.2582 | 59 | 79 | SM | 3.4 | 440 | - | 0 | 2030 | 12c | 52 | 58 | 5 |
| 246 | 2 | 1229 | 55 351480 8146450 | 0.2145 | 59 | 79 | SM | 4.1 | 440 | - | 0 | 2030 | 12c | 52 | 58 | 5 |
| 246 | 3 | 1229 | 55 351750 8146450 | 0.2307 | 59 | 79 | SM | 4.4 | 440 | W | 10 | 2030 | 12c | 52 | 58 | 5 |
| 246 | 4 | 1229 | 55 351750 8146540 | 0.1959 | 59 | 79 | SM | 7.1 | 440 | W | 10 | 2030 | 12c | 52 | 58 | 5 |
| 250 | 1 | 1229 | 55 352220 8145720 | 0.1214 | 60 | 75 | SM | 4.1 | 430 | ESE | 5 | 2030 | 2a | 49 | 59,61,70 | 5 |
| 250 | 2 | 1229 | 55 352300 8145580 | 0.1012 | 60 | 75 | SM | 4.7 | 430 | N | S | 2030 | 12c | 49 | 59,61,70 | 5 |
| 282 | 1 | 194 | 55 331950 8084360 | 0.1068 | 61 | 74 | AV | 7.6 | 1040 | W | 15 | 1650 | 9 | 60 | 60,61 | 5 |
| 282 | 2 | 194 | 55 331950 8084460 | 0.1166 | 61 | 74 | AV | - | 1040 | W | 15 | 1650 | 9 | 60 | 60,61 | 5 |
| 282 | 3 | 194 | 55 332040 8084460 | 0.1216 | 61 | 70 | AV | - | 1040 | W | 15 | 1650 | 9 | 60 | 60,61 | 5 |
| 283 | 1 | 194 | 55 332750 8089530 | 0.1445 | 61 | 74 | CG | 9.6 | 1040 | W | 5 | 1650 | 9 | 57 | 57,61 | 5 |
| 283 | 2 | 194 | 55 332750 8089550 | 0.1538 | 61 | 74 | CG | 6.9 | 1040 | W | 5 | 1650 | 9 | 57 | 57,61 | 5 |
| 283 | 3 | 194 | 55 332750 8089590 | 0.1194 | 61 | 70 | CG | 7.4 | 1040 | W | 5 | 1650 | 9 | 57 | 57,61 | 5 |
| 310 | 1 | 310 | 55 358300 8090150 | 0.0911 | 55 | 75 | BV | 4.1 | 670 | W | 5 | 2030 | 1b | 55 | 55,65 | 5 |
| 311 | 1 | 194 | 55 332200 8084450 | 0.1012 | 61 | 87 | AV | 9.2 | 1040 | SW | 15 | 1650 | 16c | 60 | 60 | 5 |
| 317 | 1 | 185 | 55 349950 8101010 | 0.1012 | 62 | 67 | SM | - | 730 | - | 0 | 1320 | | 43,51 | 62 | 5 |
| 321 | 1 | 185 | 55 354030 8105410 | 0.1012 | 61 | 71 | SM | 5.0 | 730 | NE | 10 | 1650 | | 45,60 | 60 | 5 |
| 322 | 1 | 1229 | 55 351060 8145760 | 0.1012 | 61 | 79 | SM | 7.2 | 488 | WNW | 15 | 2030 | 2a | 56 | 61,75 | 5 |
| 324 | 1 | 1229 | 55 349920 8146860 | 0.0777 | 63 | 74 | SM | 5.4 | 460 | E | 5 | 2100 | | 48 | 62 | 5 |
| 329 | 1 | 1137 | 55 400400 8026150 | 0.1590 | 63 | 82 | SM | 6.6 | 30 | - | 0 | 4000 | 2a | 60 | 62,65 | 5 |
| 329 | 2 | 1137 | 55 400450 8026250 | 0.1348 | 63 | 82 | SM | - | 30 | SW | 15 | 4000 | 2a | 60 | 62,65 | 5 |
| 331' | 1 | 185 | 55 352940 8105580 | 0.1012 | 61 | 71 | CG | - | 730 | SE | 15 | 1650 | | 58 | 62 | 5 |
| 332 | 1 | 1229 | 55 351630 8144880 | 0.1012 | 62 | 79 | SM | 7.0 | 560 | W | 10 | 2080 | | 57 | 62,74 | 5 |
| 333 | 1 | 310 | 55 360510 8089250 | 0.1064 | 61 | 78 | SM | 2.5 | 670 | SW | 10 | 2090 | | 58 | 61,73 | 5 |

| | | | | | | | | | | | | | | | | | |
|---|---|---|---|---|---|---|---|---|---|---|---|---|---|---|---|---|---|
| 347 | 1 | 310 | 55 358110 8089430 | 0.1012 | 59 | 74 | BV | 7.2 | 670 | WNW | 5 | 2000 | | 55 | 55,59 | 5 |
| 350 | 1 | 458 | 55 351100 8024380 | 0.2015 | 65 | 66 | CG | - | 600 | | | 1500 | | 64 | 65 | 5 |
| 352 | 1 | 185 | 55 352590 8101460 | 0.2764 | 65 | 75 | SM | 7.8 | 680 | SE | 5 | 1320 | | 30 | 65,71 | 5 |
| 370 | 1 | 605 | 55 351150 8024300 | 0.2023 | 69 | 84 | TG | 5.0 | 760 | NE | 5 | 2000 | 8 | 52 | | 1 |
| 370 | 2 | 605 | 55 351200 8024300 | 0.2023 | 69 | 80 | TG | 2.5 | 760 | NE | 5 | 2000 | 8 | 52 | 65,68 | 1 |
| 380 | 1 | 1229 | 55 351300 8146920 | 0.0405 | 66 | 84 | SM | 0.6 | 440 | SE | 5 | 2030 | | 52,77 | 54 | 7 |
| 380 | 2 | 1229 | 55 351280 8146910 | 0.0405 | 66 | 84 | SM | 4.4 | 440 | SE | 5 | 2030 | | 52,77 | 54 | 7 |
| 380 | 3 | 1229 | 55 351360 8146850 | 0.0405 | 66 | 84 | SM | 4.1 | 440 | SE | 5 | 2030 | | 52,77 | 54 | 7 |
| 380 | 4 | 1229 | 55 351360 8146790 | 0.0405 | 66 | 84 | SM | 3.2 | 440 | SE | 5 | 2030 | | 52,77 | 54,68 | 7 |
| 380 | 5 | 1229 | 55 351300 8146830 | 0.0405 | 66 | 84 | SM | 1.6 | 440 | SE | 5 | 2030 | | 52,77 | 54,68 | 7 |
| 380 | 6 | 1229 | 55 351240 8146870 | 0.0405 | 66 | 84 | SM | 2.5 | 440 | SE | 5 | 2030 | | 52,77 | 54,68 | 7 |
| 380 | 7 | 1229 | 55 351290 8146770 | 0.0405 | 66 | 84 | SM | 4.5 | 440 | SE | 5 | 2030 | | 52,77 | 54,68 | 7 |
| 380 | 8 | 1229 | 55 351180 8146770 | 0.0405 | 66 | 84 | SM | 0.6 | 440 | SE | 5 | 2030 | | 52,77 | 54,68 | 7 |
| 380 | 9 | 1229 | 55 351230 8146740 | 0.0405 | 66 | 84 | SM | 2.0 | 440 | SE | 5 | 2030 | | 52,77 | 54,68 | 7 |
| 380 | 10 | 1229 | 55 351360 8146720 | 0.0405 | 66 | 84 | SM | 6.7 | 440 | SE | 5 | 2030 | | 52,77 | 54 | 7 |
| 380 | 11 | 1229 | 55 351310 8146690 | 0.0405 | 66 | 84 | SM | 5.6 | 440 | SE | 5 | 2030 | | 52,77 | 54 | 7 |
| 380 | 12 | 1229 | 55 351350 8146650 | 0.0405 | 66 | 84 | SM | 6.1 | 440 | SE | 5 | 2030 | | 52,77 | 54 | 7 |
| 380 | 13 | 1229 | 55 351340 8146570 | 0.0405 | 66 | 84 | SM | 4.8 | 440 | SE | 5 | 2030 | | 52,77 | 54,68 | 7 |
| 380 | 14 | 1229 | 55 351320 8146620 | 0.0405 | 66 | 84 | SM | 7.0 | 440 | SE | 5 | 2030 | | 52,77 | 54,68 | 7 |
| 380 | 15 | 1229 | 55 351250 8146640 | 0.0405 | 66 | 84 | SM | 5.4 | 440 | SE | 5 | 2030 | | 52,77 | 54,68 | 7 |
| 380 | 16 | 1229 | 55 351200 8146660 | 0.0405 | 66 | 84 | SM | 3.4 | 440 | SE | 5 | 2030 | | 52,77 | 54,68 | 7 |
| 380 | 17 | 1229 | 55 351180 8146720 | 0.0405 | 66 | 84 | SM | 1.4 | 440 | SE | 5 | 2030 | | 52,77 | 54,68 | 7 |
| 380 | 18 | 1229 | 55 351120 8146710 | 0.0405 | 66 | 84 | SM | 4.0 | 440 | SE | 5 | 2030 | | 52,77 | 54,68 | 7 |
| 381 | 11 | 194 | 55 332330 8085780 | 0.0405 | 66 | 84 | CG | 8.9 | 1180 | NE | 10 | 1650 | | 56,80 | 53 | 7 |
| 381 | 12 | 194 | 55 332350 8085790 | 0.0405 | 67 | 84 | CG | 6.8 | 1180 | NE | 10 | 1650 | | 56,80 | 53 | 7 |
| 381 | 17 | 194 | 55 331950 8085970 | 0.0405 | 67 | 84 | CG | 8.4 | 1180 | SW | 5 | 1650 | | 56,80 | 53,67 | 7 |
| 381 | 18 | 194 | 55 331950 8085920 | 0.0405 | 67 | 77 | CG | 5.2 | 1180 | SW | 10 | 1650 | | 56 | 53,67 | 7 |
| 381 | 21 | 194 | 55 331790 8085870 | 0.0405 | 67 | 84 | CG | 2.5 | 1180 | SE | 5 | 1650 | | 56,80 | 53,67 | 7 |
| 381 | 22 | 194 | 55 331750 8085890 | 0.0405 | 67 | 84 | CG | 6.0 | 1180 | SE | 5 | 1650 | | 56,80 | 53,67 | 7 |
| 381 | 24 | 194 | 55 331930 8086000 | 0.0405 | 67 | 84 | CG | - | 1180 | SW | 5 | 1650 | | 56,80 | 53,67 | 7 |
| 381 | 25 | 194 | 55 331910 8086000 | 0.0405 | 67 | 84 | CG | 2.5 | 1180 | SW | 10 | 1650 | | 56,80 | 53,67 | 7 |
| 381 | 26 | 194 | 55 331730 8085970 | 0.0405 | 70 | 84 | CG | - | 1180 | E | 10 | 1650 | | 56,80 | 53,67 | 7 |
| 381 | 27 | 194 | 55 331750 8085920 | 0.0405 | 67 | 84 | CG | 3.9 | 1180 | E | 5 | 1650 | | 56,80 | 53 | 7 |
| 381 | 28 | 194 | 55 331650 8085970 | 0.0405 | 67 | 84 | CG | 8.6 | 1180 | NE | 10 | 1650 | | 56,80 | 53 | 7 |
| 381 | 29 | 194 | 55 331640 8085900 | 0.0405 | 71 | 84 | CG | 4.7 | 1180 | E | 10 | 1650 | | 56,80 | 53,67 | 7 |
| 381 | 30 | 194 | 55 331640 8085930 | 0.0405 | 71 | 84 | CG | 7.2 | 1180 | E | 10 | 1650 | | 56,80 | 53,67 | 7 |
| 408 | 41 | 194 | 55 331690 8087040 | 0.2023 | 69 | 75 | CG | 6.2 | 1130 | S | 10 | 1650 | | 69 | | 4 |
| 408 | 42 | 194 | 55 331600 8086960 | 0.2023 | 69 | 75 | CG | 6.1 | 1130 | SW | 10 | 1650 | | 69 | | 4 |
| 408 | 43 | 194 | 55 331500 8087040 | 0.2023 | 69 | 75 | CG | 3.7 | 1130 | S | 10 | 1650 | | 69 | 69 | 4 |
| 408 | 44 | 194 | 55 331400 8086960 | 0.2023 | 69 | 75 | CG | 3.8 | 1130 | S | 10 | 1650 | | 69 | 69 | 4 |
| 408 | 45 | 194 | 55 331300 8087040 | 0.2023 | 69 | 75 | CG | 4.4 | 1130 | SW | 5 | 1650 | | 69 | 69 | 4 |
| 408 | 46 | 194 | 55 331210 8086960 | 0.2023 | 69 | 75 | CG | 4.8 | 1130 | SW | 5 | 1650 | | 69 | 69 | 4 |
| 408 | 47 | 194 | 55 331400 8087220 | 0.2023 | 69 | 75 | CG | 3.9 | 1130 | NE | 5 | 1650 | | 69 | 69 | 4 |
| 408 | 48 | 194 | 55 331500 8087170 | 0.2023 | 69 | 75 | CG | 4.7 | 1130 | NE | 5 | 1650 | | 69 | 69 | 4 |
| 423 | 1 | 756 | 55 354920 8052880 | 0.2023 | 68 | 76 | BV | - | 820 | NE | 20 | 2500 | | 45,64 | 68 | 5 |

| | | | | | | | | | | | | | | | | | |
|---|---|---|---|---|---|---|---|---|---|---|---|---|---|---|---|---|---|
| 423 | 2 | 756 | 55 355080 8052650 | 0.2023 | 68 | 73 | BV | - | 820 | NE | 20 | 2500 | | 45,64 | 68 | 5 |
| 423 | 3 | 756 | 55 355030 8052720 | 0.2023 | 68 | 73 | BV | 6.2 | 820 | NE | 20 | 2500 | | 45,64 | 68 | 5 |
| 423 | 4 | 756 | 55 354970 8052800 | 0.2023 | 68 | 73 | BV | - | 820 | NE | 20 | 2500 | | 45,64 | 68 | 5 |
| 423 | 5 | 756 | 55 355020 8052860 | 0.2023 | 68 | 73 | BV | - | 820 | NE | 20 | 2500 | | 45,64 | 68 | 5 |
| 423 | 6 | 756 | 55 354960 8052920 | 0.2023 | 68 | 73 | BV | - | 820 | NE | 20 | 2500 | | 45,64 | 68 | 5 |
| 423 | 7 | 756 | 55 355200 8052740 | 0.2023 | 68 | 73 | BV | - | 820 | SW | 10 | 2500 | | 45,64 | 68 | 5 |
| 423 | 8 | 756 | 55 355140 8052800 | 0.2023 | 68 | 73 | BV | 6.2 | 820 | SW | 10 | 2500 | | 45,64 | 68 | 5 |
| 423 | 9 | 756 | 55 355130 8052950 | 0.2023 | 68 | 73 | BV | 5.7 | 820 | SW | 10 | 2500 | | 45,64 | 68 | 5 |
| 423 | 10 | 756 | 55 355190 8052830 | 0.2023 | 68 | 73 | BV | 7.4 | 820 | SW | 10 | 2500 | | 45,64 | 68 | 5 |
| 423 | 11 | 756 | 55 355260 8052780 | 0.2023 | 68 | 73 | BV | - | 820 | SW | 10 | 2500 | | 45,64 | 68 | 5 |
| 423 | 12 | 756 | 55 355070 8053020 | 0.2023 | 68 | 73 | BV | 4.8 | 820 | SW | 10 | 2500 | | 45,64 | 68 | 5 |
| 431 | 1 | 1137 | 55 401150 8025700 | 0.1012 | 64 | 87 | SM | 7.2 | 30 | - | 0 | 4000 | 2a | 50 | 56,64 | 1 |
| 431 | 2 | 1137 | 55 400700 8025700 | 0.1012 | 64 | 87 | SM | - | 30 | N | 5 | 4000 | 2a | 50 | 56,64 | 1 |
| 434 | 1 | 310 | 55 361160 8086590 | 0.2023 | 70 | 83 | BV | 6.5 | 670 | W | 10 | 2000 | | 30,69 | 29 | 3 |
| 434 | 2 | 310 | 55 361160 8086630 | 0.2023 | 70 | 83 | BV | 5.8 | 670 | SE | 10 | 2000 | | 30,69 | 29,70 | 3 |
| 434 | 3 | 310 | 55 361150 8086830 | 0.2023 | 70 | 83 | BV | 5.9 | 670 | W | 10 | 2000 | | 30,69 | 29,70 | 3 |
| 434 | 4 | 310 | 55 361150 8086920 | 0.2023 | 70 | 83 | BV | 6.5 | 670 | W | 10 | 2000 | | 30,69 | 29,70 | 3 |
| 434 | 5 | 310 | 55 361270 8086670 | 0.2023 | 70 | 83 | BV | 6.1 | 670 | W | 10 | 2000 | | 30,69 | 29,70 | 3 |
| 434 | 6 | 310 | 55 361360 8086673 | 0.2023 | 70 | 83 | BV | 6.2 | 670 | W | 10 | 2000 | | 30,69 | 29,70 | 3 |
| 434 | 7 | 310 | 55 361410 8086630 | 0.2023 | 70 | 83 | BV | 6.5 | 670 | S | 10 | 2000 | | 30,69 | 29,70 | 3 |
| 434 | 8 | 310 | 55 361410 8086550 | 0.2023 | 70 | 83 | BV | 9.4 | 670 | N | 10 | 2000 | | 30,69 | 29,70 | 3 |
| 434 | 9 | 310 | 55 361550 8086550 | 0.2023 | 70 | 83 | BV | 5.8 | 670 | S | 10 | 2000 | | 30,69 | 29,70 | 3 |
| 434 | 10 | 310 | 55 361500 8086620 | 0.2023 | 70 | 83 | BV | 6.6 | 670 | S | 10 | 2000 | | 30,69 | 29 | 3 |
| 434 | 11 | 310 | 55 361630 8086590 | 0.2023 | 70 | 83 | BV | 5.6 | 670 | S | 20 | 2000 | | 30,69 | 29,70 | 3 |
| 434 | 12 | 310 | 55 361630 8086520 | 0.2023 | 70 | 83 | BV | 5.2 | 670 | SE | 20 | 2000 | | 30,69 | 29 | 3 |
| 450 | 1 | 310 | 55 366880 8079880 | 0.2023 | 70 | 84 | BV | 1.2 | 760 | NW | 5 | 2000 | | 51,70 | 70,73 | 5 |
| 450 | 2 | 310 | 55 366735 8079950 | 0.2023 | 70 | 72 | BV | 1.4 | 760 | NW | 5 | 2000 | | 51,70 | 70,73 | 5 |
| 450 | 3 | 310 | 55 366770 8079830 | 0.2023 | 70 | 84 | BV | 2.2 | 760 | NW | 5 | 2000 | | 51,70 | 70,73 | 5 |
| 450 | 4 | 310 | 55 366690 8079855 | 0.1214 | 70 | 72 | BV | - | 760 | NW | 5 | 2000 | | 51,70 | 70,73 | 5 |
| 450 | 5 | 310 | 55 366675 8079805 | 0.1214 | 70 | 72 | BV | 4.5 | 760 | NW | 5 | 2000 | | 51,70 | 70,73 | 5 |
| 450 | 6 | 310 | 55 366760 8079770 | 0.1214 | 70 | 72 | BV | 5.4 | 760 | NW | 5 | 2000 | | 51,70 | 70,73 | 5 |
| 450 | 7 | 310 | 55 366830 8079750 | 0.1214 | 70 | 72 | BV | - | 760 | NW | 5 | 2000 | | 51,70 | 70,73 | 5 |
| 456 | 2 | 194 | 55 331660 8086520 | 0.1518 | 69 | 77 | CG | 8.8 | 1060 | N | 5 | 1650 | | 54 | 54 | 5 |
| 469 | 11 | 607 | 55 353550 8126520 | 0.1619 | 70 | 82 | CG | 4.8 | 460 | S | 5 | 1800 | | 59 | 70 | 8 |
| 469 | 12 | 607 | 55 353560 8126500 | 0.1619 | 70 | 82 | CG | 4.6 | 460 | S | 5 | 1800 | | 59 | 70 | 8 |
| 469 | 13 | 607 | 55 353570 8126500 | 0.1619 | 70 | 82 | CG | 3.4 | 460 | S | 5 | 1800 | | 59 | 70 | 8 |
| 469 | 14 | 607 | 55 353540 8126520 | 0.1619 | 70 | 82 | CG | 5.1 | 460 | S | 5 | 1800 | | 59 | 70 | 8 |
| 469 | 21 | 607 | 55 353510 8126510 | 0.1619 | 70 | 82 | CG | 4.9 | 460 | W | 5 | 1800 | | 59 | 70 | 8 |
| 469 | 22 | 607 | 55 353530 8126490 | 0.1619 | 70 | 82 | CG | 6.2 | 460 | W | 5 | 1800 | | 59 | 70 | 8 |
| 469 | 23 | 607 | 55 353510 8126480 | 0.1619 | 70 | 82 | CG | 6.5 | 460 | W | 5 | 1800 | | 59 | 70 | 8 |
| 469 | 24 | 607 | 55 353540 8126500 | 0.1619 | 70 | 82 | CG | 9.1 | 460 | W | 5 | 1800 | | 59 | 70 | 8 |
| 469 | 31 | 607 | 55 353530 8126530 | 0.1619 | 70 | 82 | CG | 1.9 | 460 | SW | 5 | 1800 | | 59 | 70 | 8 |
| 469 | 32 | 607 | 55 353540 8126510 | 0.1619 | 70 | 82 | CG | 5.2 | 460 | SW | 5 | 1800 | | 59 | 70 | 8 |
| 469 | 33 | 607 | 55 353550 8126530 | 0.1619 | 70 | 82 | CG | 6.1 | 460 | SW | 5 | 1800 | | 59 | 70 | 8 |
| 469 | 34 | 607 | 55 353560 8126510 | 0.1619 | 70 | 82 | CG | 6.2 | 460 | SW | 5 | 1800 | | 59 | 70 | 8 |
| 576 | 1 | 756 | 55 362000 8038200 | PRISM | 78 | 87 | BV | - | 760 | SW | 10 | 2500 | | 77 | | 9 |
| 577 | 1 | 144 | 55 294450 8198710 | PRISM | 77 | 86 | CG | - | 1006 | E | 18 | 1036 | | 77 | | 9 |
| 582 | 1 | 144 | 55 292680 8200130 | PRISM | 77 | 87 | CG | - | 1070 | S | 13 | 1970 | | 77 | | 9 |
| 591 | 1 | 607 | 55 353600 8115540 | 0.4087 | 52 | 84 | CG | 8.2 | 730 | SW | 15 | 2200 | 8/9 | | | 10 |
| 594 | 1 | 310 | 55 357900 8089980 | 0.0741 | 51 | 83 | BV | 7.0 | 720 | NE | 5 | 2030 | 1b | | | 10 |
| 594 | 2 | 310 | 55 357900 8090000 | 0.1437 | 51 | 83 | BV | 8.3 | 720 | NE | 5 | 2030 | 1b | | | 10 |
| 594 | 3 | 310 | 55 357900 8090020 | 0.0660 | 51 | 83 | BV | 7.1 | 720 | NE | 5 | 2030 | 1b | | | 10 |
| 595 | 1 | 310 | 55 361020 8086910 | 0.3237 | 51 | 83 | BV | 5.9 | 670 | SW | 10 | 2000 | | 29 | 29 | 10 |
| 598 | 1 | 755 | 55 358550 8073100 | PRISM | 78 | 85 | BV | - | 520 | SW | 5 | 2540 | | 62 | | 9 |
| 606 | 1 | 185 | 55 357120 8098500 | 0.1012 | 50 | 83 | BV | 7.5 | 720 | SW | 5 | 1850 | | 16 | 53,65 | 11 |
| 608 | 1 | 310 | 55 364300 8078600 | 0.2023 | 51 | 87 | BV | 7.4 | 760 | N | 5 | 2290 | 1b | 49 | | 4 |
| 608 | 2 | 310 | 55 364300 8078600 | 0.1955 | 53 | 87 | BV | 6.9 | 760 | E | 5 | 2290 | 1b | 49,72 | 53,66 | 4 |
| 608 | 3 | 310 | 55 364300 8078600 | 0.2064 | 56 | 87 | BV | 6.4 | 760 | E | S | 2290 | | 49 | 56,66 | 4 |
| 609 | 1 | 2S1 | 55 343480 8041060 | 0.1S18 | 51 | 87 | AV | S.6 | 770 | S | 10 | 1700 | 9 | 50 | | 4 |
| 609 | 2 | 2S1 | 55 343520 8041040 | 0.2003 | 52 | 87 | AV | 3.8 | 770 | S | 20 | 1700 | 9 | 50 | 52 | 4 |
| 610 | 1 | 1229 | 55 352380 8145710 | 0.2023 | 51 | 87 | SM | 7.3 | 440 | E | S | 2030 | 12c | 49 | | 4 |
| 610 | 2 | 1229 | 55 352340 8145750 | 0.2023 | 52 | 87 | SM | S.8 | 440 | E | S | 2030 | 12c | 49 | 52,66 | 4 |
| 610 | 3 | 1229 | 55 352320 8145780 | 0.2023 | 55 | 87 | SM | 6.0 | 440 | NE | S | 2030 | | 49 | 55 | 4 |
| 611 | 1 | VCL | 55 376200 8022500 | 0.2023 | 51 | 68 | AL | 3.4 | 20 | N | S | 3800 | la | 50 | | 4 |
| 611 | 2 | VCL | 55 376250 8022500 | 0.2064 | 52 | 68 | AL | S.2 | 20 | N | S | 3800 | la | 50 | 52,58 | 4 |
| 611 | 3 | VCL | 55 375700 8022500 | 0.2023 | 55 | 68 | AL | 6.4 | 20 | N | S | 3800 | la | 50 | 55 | 4 |
| 612 | 1 | 268 | 55 409800 7904700 | 0.2023 | 51 | 86 | CG | 7.3 | 550 | N | S | 1900 | 8/6 | 51 | | 4 |
| 612 | 2 | 268 | 55 409700 7904600 | 0.2023 | 52 | 86 | CG | S.1 | 550 | N | S | 1900 | | 51 | 52,66 | 4 |
| 612 | 3 | 268 | 55 409800 7904700 | 0.2023 | 55 | 86 | CG | 7.1 | 550 | N | S | 1900 | | 51 | 55,73 | 4 |
| 613 | 1 | 344 | 55 367550 7987550 | 0.2023 | 51 | 86 | CG | 6.8 | 600 | E | S | 1300 | 2a | 47 | | 4 |
| 613 | 2 | 344 | 55 367530 7987570 | 0.2023 | 52 | 86 | CG | 3.3 | 600 | E | S | 1300 | 2a | 47 | 52,66 | 4 |
| 613 | 3 | 344 | 55 367580 7987590 | 0.2023 | 55 | 86 | CG | 7.2 | 600 | E | S | 1300 | | 47 | 55,66 | 4 |
| 614 | 1 | 1137 | 55 401650 8020500 | 0.2193 | 52 | 87 | SM | 3.8 | 30 | E | 1S | 4000 | 2a | 50 | | 4 |
| 614 | 2 | 1137 | 55 401650 8020620 | 0.2193 | 52 | 87 | SM | 5.0 | 30 | E | 20 | 4000 | | 50 | 52 | 4 |
| 614 | 3 | 1137 | 55 401560 8020410 | 0.2023 | 55 | 87 | SM | S.6 | 30 | SSW | 20 | 4000 | | 50 | 55,66 | 4 |
| 61S | 1 | 194 | 55 335650 8079850 | 0.2023 | 52 | 85 | CG | 7.7 | 1100 | E | 20 | 1650 | | 47,54,77 | 52,66 | 4 |
| 61S | 2 | 194 | 55 335600 8079900 | 0.2024 | 52 | 85 | CG | 8.4 | 1100 | E | 20 | 1650 | | 47,54,77 | 52,66 | 4 |
| 61S | 3 | 194 | 55 335100 8079600 | 0.2185 | 52 | 85 | CG | 8.4 | 1100 | W | 10 | 1650 | 14a | 47,54 | | 4 |
| 61S | 4 | 194 | 55 334700 8079200 | 0.2023 | 52 | 85 | CG | 7.2 | 1100 | S | S | 1650 | | 47 | 52,66 | 4 |
| 61S | S | 194 | 55 334600 8079300 | 0.2023 | 52 | 85 | CG | 9.1 | 1100 | S | S | 1650 | | 47,54 | 52,66 | 4 |
| 61S | 6 | 194 | 55 334650 8079200 | 0.2064 | 52 | 85 | CG | 9.8 | 1100 | S | S | 1650 | 14a | 47,54 | | 4 |
| 616 | 1 | 194 | 55 332050 8086220 | 0.2084 | 52 | 87 | CG | 9.2 | 1100 | NE | S | 1650 | 9 | 51,80 | | 4 |
| 616 | 2 | 194 | 55 332070 8086280 | 0.2023 | 52 | 87 | CG | 7.1 | 1100 | NE | S | 1650 | 9 | 51,80 | 52,66 | 4 |
| 616 | 3 | 194 | 55 332050 8086280 | 0.2023 | 52 | 87 | CG | 6.8 | 1100 | NE | S | 1650 | 9 | 51,80 | 52,66 | 4 |
| 617 | 1 | 194 | 55 331850 8086100 | 0.1012 | 53 | 87 | CG | 7.6 | 1130 | E | 10 | 1650 | | 51,80 | 53 | 4 |
| 617 | 2 | 194 | 55 331800 8086150 | 0.1012 | 53 | 87 | CG | 7.7 | 1130 | E | 10 | 1650 | | 51,80 | 53,60 | 4 |
| 617 | 3 | 194 | 55 331700 8086150 | 0.1016 | 53 | 87 | CG | S.7 | 1130 | - | 0 | 1650 | | 51,80 | 53,60 | 4 |
| 617 | 4 | 194 | 55 331700 8086100 | 0.1016 | 53 | 87 | CG | 8.3 | 1130 | E | 20 | 1650 | | 51,80 | 53,60 | 4 |
| 617 | S | 194 | 55 331750 8086100 | 0.1028 | 53 | 87 | CG | 6.6 | 1130 | E | 20 | 1650 | | 51,80 | 53,60 | 4 |
| 617 | 6 | 194 | 55 331900 8086050 | 0.1416 | 53 | 87 | CG | 3.9 | 1130 | NE | S | 1650 | | 51,80 | 52,60,66 | 4 |
| 618 | 1 | 2S1 | 55 348080 8038100 | 0.2003 | 54 | 87 | BV | 9.1 | 740 | - | 0 | 1800 | 5a | 52,67 | | 4 |
| 618 | 2 | 2S1 | 55 348060 8037340 | 0.2023 | 56 | 87 | BV | 6.2 | 760 | SE | S | 1800 | 5a | 52,67 | 56 | 4 |
| 618 | 3 | 2S1 | 55 348080 8038050 | 0.2023 | 56 | 87 | BV | 6.5 | 740 | - | 0 | 1800 | | 52,67 | 56 | 4 |
| 619 | 1 | 4S8 | 55 375500 7928410 | 0.3966 | 54 | 86 | CG | 8.7 | 600 | NW | S | 1500 | 8 | 47,73 | | 4 |
| 619 | 2 | 4S8 | 55 375500 7928350 | 0.4047 | 54 | 66 | CG | 7.7 | 600 | NW | 5 | 1500 | 8/6 | 47 | 54 | 4 |
| 619 | 3 | 458 | 55 375550 7928350 | 0.1619 | 66 | 86 | CG | 4.9 | 600 | NW | 5 | 1500 | 8/6 | 47,73 | 54,66 | 4 |
| 619 | 4 | 4S8 | 55 375450 7928580 | 0.1619 | 66 | 86 | CG | S.6 | 600 | NW | S | 1500 | | 47,73 | 54,66 | 4 |
| 620 | 1 | 1229 | 55 351180 8146840 | 0.4047 | 55 | 87 | SM | 3.0 | 440 | E | 10 | 2030 | 12c | 52,76 | 55,76 | 4 |
| 621 | 1 | 194 | 55 332710 8087400 | 0.2023 | 68 | 85 | CG | 7.4 | 1100 | W | 1S | 1650 | 9 | 64 | | 4 |
| 621 | 2 | 194 | 55 332450 8087360 | 0.2023 | 68 | 85 | CG | 4.7 | 1100 | SE | 1S | 1650 | 9 | 64 | 68 | 4 |
| 622 | 1 | 310 | 55 360500 8091400 | 0.2023 | 68 | 85 | BV | 6.0 | 640 | NW | S | 2030 | 1b | 67 | | 4 |
| 622 | 2 | 310 | 55 360620 8091400 | 0.2023 | 68 | 85 | BV | 6.6 | 640 | SE | 10 | 2030 | 1b | 67 | 68 | 4 |
| 623 | 1 | 1229 | 55 349600 8148520 | 0.2023 | 68 | 84 | SM | 6.1 | 430 | NW | S | 2030 | 2a | 57 | | 4 |
| 623 | 2 | 1229 | 55 349680 8148550 | 0.2023 | 68 | 84 | SM | 7.4 | 430 | NW | S | 2030 | 2a | 57 | 68 | 4 |
| 623 | 3 | 1229 | 55 349660 8148580 | 0.2023 | 68 | 84 | SM | 7.S | 430 | NW | S | 2030 | 2a | 57 | 68 | 4 |
| 624 | 1 | 60S | 55 352200 8025050 | 0.2023 | 68 | 84 | TG | S.8 | 760 | NE | 20 | 2000 | 8 | 51 | | 11 |
| 624 | 2 | 60S | 55 351980 8024650 | 0.2023 | 68 | 84 | TG | 6.6 | 760 | NE | S | 2000 | 8 | 52 | | 11 |
| 624 | 3 | 60S | 55 352980 8024550 | 0.2023 | 68 | 84 | TG | 4.7 | 760 | SW | 1S | 2000 | 8 | 51 | | 11 |
| 624 | 4 | 60S | 55 352680 8024180 | 0.2023 | 68 | 84 | TG | 7.2 | 760 | W | 2S | 2000 | 8 | 51 | | 11 |
| 624 | S | 60S | 55 349850 8024060 | 0.2023 | 68 | 84 | TG | S.8 | 760 | SW | S | 2000 | 8 | 52,80 | | 11 |
| 62S | 1 | 18S | 55 354400 8106740 | 0.2023 | 68 | 84 | CG | 9.6 | 700 | W | 1S | 1650 | 8 | 52 | | 11 |
| 62S | 2 | 18S | 55 350510 8107810 | 0.2023 | 69 | 84 | CG | 8.8 | 94S | S | 2S | 1650 | 8/9 | | | 11 |
| 62S | 3 | 185 | 55 352310 8108900 | 0.2023 | 68 | 84 | CG | 8.8 | 1065 | SW | 2S | 1650 | 9 | 65 | | 11 |
| 62S | 4 | 18S | 55 351200 8107360 | 0.2023 | 68 | 84 | CG | 6.9 | 790 | E | 1S | 1650 | 8/9 | 52 | | 11 |
| 62S | S | 185 | 55 351150 8106620 | 0.2023 | 68 | 84 | CG | 7.2 | 730 | SE | 20 | 1650 | 8 | 70 | | 11 |
| 626 | 1 | 1229 | 55 355500 8143090 | 0.2023 | 69 | 87 | SM | S.4 | 360 | SSW | S | 2030 | 2a | 54 | | 11 |
| 626 | 2 | 1229 | 55 354760 8143540 | 0.2023 | 69 | 87 | SM | 7.0 | 360 | NE | 10 | 2030 | 2a | | | 11 |
| 640 | 1 | 7S6 | 55 361100 8052500 | PRISM | 79 | 86 | BV | - | 720 | N | 15 | 2000 | | 62,80 | | 9 |
| 679 | 1 | 144 | 55 290900 8201600 | PRISM | 80 | 85 | CG | - | 1100 | N | S | 1036 | | 80 | | 9 |
| 679 | 2 | 144 | 55 290600 8201700 | PRISM | 80 | 85 | CG | - | 1060 | NE | S | 1036 | | | | 9 |
| 701 | 1 | 756 | 55 371650 8047300 | PRISM | 85 | 88 | CG | - | 400 | | | 2000 | | 87 | | 9 |
| EP | 2 | 185 | 55 349550 8103510 | 0.5000 | 71 | 87 | CG | - | 720 | SE | 5 | 1200 | | 43 | | 12 |
| EP | 3 | 607 | 55 350290 8110090 | 0.5000 | 71 | 87 | CG | - | 1120 | NE | 15 | 2400 | | | | 12 |
| EP | 4 | 933 | 55 337490 8129060 | 0.5000 | 72 | 88 | CG | 6.4 | 80 | SW | 5 | 2500 | | 59 | | 12 |
| EP | 9 | 185 | 55 354440 8106960 | 0.5000 | 72 | 88 | CG | - | 710 | E | 20 | 1650 | | 63 | | 12 |
| EP | 18 | 143 | 55 311100 8169830 | 0.5000 | 73 | 87 | CG | - | 1100 | W | 5 | 2500 | | | | 12 |
| EP | 19 | 750 | 55 368800 7954780 | 0.5000 | 75 | 87 | CG | - | 620 | SE | 10 | 2000 | | | | 12 |
| EP | 29 | 650 | 55 345710 8059260 | 0.5000 | 75 | 87 | AV | - | 1200 | SE | 15 | 2700 | | | | 12 |
| EP | 30 | 144 | 55 293260 8199380 | 0.5000 | 76 | 88 | CG | - | 980 | W | 5 | 1500 | | | | 12 |
| EP | 31 | 755 | 55 375510 8061530 | 0.5000 | 76 | 88 | SM | 6.0 | 80 | S | 5 | 4000 | | | | 12 |
| EP | 32 | TR 14 | 54 752330 8479700 | 0.5000 | 75 | 87 | SM | 2.8 | 450 | SW | 5 | 2000 | | | | 12 |
| EP | 33 | 452 | 55 348100 8088250 | 0.5000 | 76 | 88 | BV | - | 720 | - | 0 | 1400 | | 52 | | 12 |
| EP | 34 | 755 | 55 369440 8074860 | 0.5000 | 76 | 88 | AL | - | 380 | SW | 5 | 4000 | | | | 12 |
| EP | 35 | TR 55 | 55 322210 8190920 | 0.5000 | 77 | 87 | SM | - | 230 | SE | 10 | 2900 | | | | 12 |
| EP | 37 | 679 | 55 660835 7649387 | 0.5000 | 77 | 87 | BV | - | 920 | SE | 5 | 2400 | | | | 12 |

| EP | 38 | 194 | 55 338220 8073640 | 0.5000 | 77 | 87 | AV | - | 1000 | SE | 10 | 1800 | 12 |
| --- | --- | --- | --- | --- | --- | --- | --- | --- | --- | --- | --- | --- | --- |
| EP | 40 | 144 | 55 297320 8198970 | 0.5000 | 78 | 88 | CG | - | 800 | N | 10 | 1300 | 12 |
| EP | 41 | NP | 55 333200 8215260 | 0.5000 | 77 | 87 | AL | - | 15 | SE | 5 | 3500 | 12 |
| EP | 42 | CL | 54 745020 8590560 | 0.5000 | 77 | 87 | AL | - | 30 | SE | 10 | 2200 | 12 |
| EP | 43 | 194 | 55 333560 8085620 | 0.5000 | 78 | 88 | AV | - | 1120 | S | 20 | 2000 | 12 |
| EP | 44 | 144 | 55 295120 8205880 | 0.5000 | 80 | 88 | CG | - | 880 | NW | 5 | 2500 | 12 |

Plot Types:
1. Paired treatment plots comparing growth with and without silvicultural treatment.
2. Plots monitoring the development of regeneration.
3. Experiments monitoring development of enrichment plantings following thinning to various spacings.
4. Experiments monitoring development of rainforest following application of different silvicultural treatment prescriptions. *5.* Experiments monitoring development of enrichment plantings.
6. Experiment examining benefits of silvicultural treatment 10 years prior to logging, with a view to getting more regeneration. 7. Experiments monitoring effects of re-treatment 15 years after initial silvicultural treatment. 8. Treatment of unproductive rainforest attempting to produce a viable timber harvest.
9. Logging damage studies.
10. Plots monitoring development of dense stands of rainforest.
11. Plots monitoring growth and yield in rainforest under routine management. These plots were deliberately located to sample good, average and poor rainforest.
12. CSIRO growth monitoring plots described in West *et al.* (1988).